\documentstyle[twocolumn,prl,aps,epsf]{revtex}

\begin{document}

\draft

\title{Excitation spectrum in a cylindrical Bose-Einstein gas}

\author{ Tomoya Isoshima$^{\ast}$, and  Kazushige Machida}
\address{Department of Physics, Okayama University,
         Okayama 700, Japan}
\date{December 9, 1997}

\maketitle

\begin{abstract}
Whole excitation spectrum is calculated within the
Popov approximation of the Bogoliubov theory for a cylindrical symmetric
Bose-Einstein gas trapped radially by a harmonic potential.
The full dispersion relation and its temperature dependence
of the zero sound mode propagating along the axial direction
are evaluated in a self-consistent manner.
The sound velocity is shown to depend not only on the peak density,
but also on the axial area density.
Recent sound velocity experiment on Na atom gas is discussed in light of
the present theory.
\end{abstract}


Much attention has been focused on Bose-Einstein 
condensation (BEC) since its realization in alkaline atomic gases.
This macroscopic quantum phenomenon is unique because it occurs in
ideal Bose gas, but real interest lies
in the fact of how the particle interaction affects on it.
Superfluidity which is absent in ideal Bose gas is understood
to be sustained by the particle interaction.
By flurrious experimental and theoretical studies,
various fundamental properties of Bose-Einstein condensation,
including the condensation fraction relative to non-condensate,
 its temperature dependence and
few lowest-lying collective modes have been already elucidated
quite thoroughly\cite{review}.
However, because of the small limited size of
trapped atomic gases these collective modes are all localized so far,
thus the spectrum was discrete\cite{localized}.

Rapid development of experimental techniques to
trap larger number of atoms leads to a chance to investigate
propagating collective modes, such as zero, first and second sound.
In particular, the zero sound mode with the continuous spectrum is
intrigue because it gives the critical superfluid
velocity above which the system returns to be normal gas.
In quantum liquid of superfluid $^4$He where phonon,
roton and maxon are known to exist microscopic description for
these modes has not been quite successful because $^4$He
atoms are strongly interacting.
In contrast, in the present dilute Bose gases there is a
good opportunity to thoroughly understand these
low-lying excitations in a microscopic level.
In that context a recent zero sound measurement
by Andrews {\it et al.}\cite{andrews}
on a long cigar shaped Na atomic gas with length typically
500$\mu {\rm m}$ and width 10$\mu {\rm m}$ provides an interesting
testing ground and challenges theoretical investigation.

The speed of this compressional zero sound mode
is known to be given by
$c=\sqrt {n_0g/m}$
with 
the repulsive interaction constant $g$ in the case of a
homogeneous bulk system where the lowest collective mode
is exhausted by the zero sound mode\cite{nozieres}.
Here $n_0$ is the condensate density.
It is not obvious at all that this expression by regarding $n_0$
as the peak density is readily applicable to the
above experiment as done by Andrews {\it et al.}\cite{andrews}
because the system trapped by a harmonic potential
is finite and everywhere inhomogeneous.

In fact, Zaremba\cite{zaremba} derives a different formula
$c=\sqrt {n_0(0)g/2m}$ for an idealized cylindrical trap
by solving a time-dependent Gross-Pitaevskii equation within the
Thomas-Fermi approximation (TFA)
with $n_0(0)$ being the peak condensate density.
The extra factor $\sqrt{2}$ in the denominator comes
from a simple geometrical reason that the average condensate
density is half the peak density in this parabolic shaped condensate.
The same result is also obtained by Kavoulakis and Pethick\cite{pethick}.
Zaremba's expression is far off the experimental data.
As pointed out by himself a microscopic calculation is needed.

Here we study the whole excitation spectrum, not only low-lying excitations,
but also high energy ones for a cylindrical BEC system which
is confined radially by a harmonic potential and infinite axially,
approximating the above long cigar shaped trapped gas.
We give special attention on zero sound velocity.
The present set of the microscopic results,
including condensate and non-condensate spatial profile,
the relative fraction, 
their $T$-dependences and the whole excitation spectrum,
within Popov approximation of the Bogoliubov
theory enables us to draw a self-consistent picture of BEC systems.
This theoretical framework has been quite successfully applied to
the present dilute atomic BEC systems so far\cite{review}.

We start out with the following Hamiltonian in which 
Bose particles interact with a two-body potential
$g\delta({\bf r}_1 - {\bf r}_2)$
where $g$ is a positive (repulsive) constant
proportional to the s-wave scattering length $a$,
namely $g=4\pi \hbar^2a/ m$:
\begin{equation}
\hat{{\rm H}}=\int d{\bf r}
   \hat{\Psi}^{\dagger} h({\bf r})\hat{\Psi}
   + \frac{g}{2} 
   \int d{\bf r}
   \hat{\Psi}^{\dagger} \hat{\Psi}^{\dagger} \hat{\Psi} \hat{\Psi}
\end{equation}
where the one-body Hamiltonian
$h({\bf r}) = -\frac{\hbar^2\nabla^2}{2m} + V({\bf r}) - \mu $,
the chemical potential $\mu$ is introduced
to fix the total particle number
and $V({\bf r})$ is the confined potential.
In order to describe the Bose condensation,
we assume that the field operator
$\hat{\Psi}$ is decomposed into
$ \hat{\Psi} = \hat{\psi} + \phi $
where the ground state average is given by
$\langle \hat{\Psi} \rangle = \phi.$
A c-number $\phi$ corresponds to the condensate wave function
and $\hat{\psi}$ is a q-number describing the non-condensate.
Thus the condensate density is given by 
$n_0=|\phi|^2$.
The condition that the first order term in
$\hat{\psi}$ vanish yields
\begin{equation}
   h({\bf r})\phi + g|\phi|^2\phi + 2g\rho\phi = 0.
\end{equation}
When the non-condensate density $\rho$ is made to zero,
it reduces to the Gross-Pitaevskii (GP) equation:
$h({\bf r})\phi + g|\phi|^2\phi=0$.
The condition that the Hamiltonian be diagonalized
gives rise to the following set
of eigenvalue equations for $u_q$ and $v_q$ with 
the eigenvalue $\varepsilon_q$:
\begin{eqnarray}
\{h({\bf r}) + 2g\rho + 2g|\phi|^2\}u_q - g \phi ^2v_q
  &=&  \varepsilon_qu_q
\\
\{h({\bf r}) + 2g\rho + 2g|\phi|^2\}v_q - g\phi ^{*2} u_q
  &=& -\varepsilon_qv_q.
\end{eqnarray}
The variational parameter
$\rho$ is determined self-consistently by
$\rho=\langle \hat{\psi}^\dagger \hat{\psi} \rangle 
=\sum_{\it q}(u_q^*u_q + v_q^*v_q)f(\varepsilon_q)
+ \sum_{\it q} v_q^*v_q$
where $f(\varepsilon)$ is the Bose distribution function.
These constitute a complete set of the equations
within the Popov approximation.
The total particle number density is
given as $n_t= n_0 + \rho$.

We now consider a cylindrically symmetric system
in the cylindrical coordinate:  ${\bf r} = (r,\theta ,z)$.
The system is trapped by a harmonic potential
$V(r)={1\over 2}m\omega_0 r^2$ radially and 
periodic along the $z$-axis whose period is $L$.
The eigenfunctions of $u_q$  and  $v_q$ are written as 
$u_q = u_q(r)e^{il\theta}e^{ikz}$ and $v_q = v_q(r)e^{il\theta}e^{ikz}$.
The set of the quantum numbers $q$ is
described by $(\kappa, l, k)$ where
$\kappa = 0,1, 2,\cdots $, 
$l= 0, \pm 1, \pm 2, \cdots$, 
$k = 0, \pm 2\pi /L, \pm 4\pi /L, \cdots$.
Then, the functions
$u_q(r)$  and $v_q(r)$
are expanded in terms of
$\varphi_l^{(i)}(r)=
\frac{\sqrt{2}}{J_{l +1}
\left(\alpha_l^{(i)}\right)}J_l\left(\alpha_l^{(i)}\frac{r}{R}\right)$
as
$ u_q(r) = \sum_i c_q^{(i)}\varphi_l^{(i)}(r)$
 and
$ v_q(r) = \sum_i d_q^{(i)}\varphi_l^{(i)}(r) $
where $J_l(r)$ is the Bessel function of $l$-th order,
$\alpha_l^{(i)}$ denotes $i$-th zero of $J_l$,
and $R$ is a sufficiently large radius.
The eigenvalue problem reduces to diagonalizing a finite matrix.
Note that since the angular momentum $l$ and
the wave number $k$ are both good quantum numbers,
the eigenvalue equation is decomposed into each $l$ and $k$.
Each block-diagonal eigenvalue equation
gives rise to the quantum number $\kappa$ along the radial direction.
The iterative calculations yield a converged self-consistent
solution of Eqs. (2), (3) and (4)\cite{isoshima}.

The zero sound experiment by Andrews {\it et al.}\cite{andrews} is done
under the fixed total number of Na atoms ($\sim5\times10^6$)
and fixed axial trapping frequency ($\sim18{\rm Hz}$)
by varying the radial trapping
frequency $\omega_0$ (typically $\sim200{\rm Hz}$),
giving rise to the zero sound speed as a function of
the peak condensate density $n_0(0)$.
To simulate this experiment we have performed self-consistent calculations
for various $\omega_0$ values
and various area density $n_z$ per unit length  
along the $z$-axis,
which is uniform in our setting
(for example, $n_z = 2 \times 10^{4} /\mu {\rm m}$).
The scattering strength is chosen as $a=2.75{\rm nm}$\cite{andrews}.
The area density $n_z$ 
turns out to be one of the key parameters
to fully characterize the experimental results,
in particular, the zero sound velocity.
The energy is scaled by $\omega_0$.
\begin{figure}
\epsfxsize=5.5cm
\hspace{0.75cm}
\epsfbox{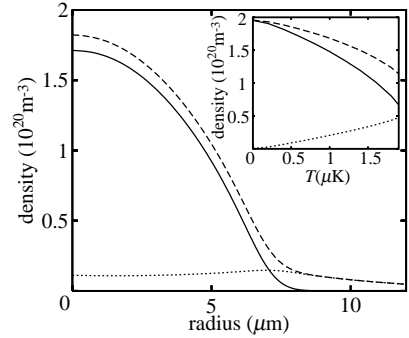}
\caption{
Spatial profiles of the condensate $n_0(r)$ (full line),
non-condensate $\rho(r)$ (dotted
line) and total $n_t(r)$ (dashed line) along the radial direction.
$n_z=2\times10^4/\mu {\rm m}$,
$\omega_0=200{\rm Hz}$ and $T=0.6\mu {\rm K}$.
The inset shows the $T$-dependences of $n_0(0)$, $\rho(0)$ and $n_t(0)$. 
}
\label{n0:rho}
\end{figure}

In Fig.~\ref{n0:rho} we show a typical example of the spatial profiles of
condensate $n_0(r)$, the non-condensate $\rho(r)$ and
the total  $n_t(r)$ along the radial direction at a finite temperature.
The spatial extension of the gas is an order of 10$\mu {\rm m}$,
which coincides with the typical radial width of a long
cigar in the experiment\cite{andrews}.
It is also seen that the condensate concentrates at the center $r=0$,
whose spatial variation is quite parabolic
as predicted by TFA\cite{zaremba} while the non-condensate
which is absent in TFA is pushed outward.
The total density profile $n_t(r)$ deviates far from a parabolic shape.
The peak condensate density $n_0(0)$ decreases progressively
as $T$ increases as shown in inset of Fig.~\ref{n0:rho}.
These results are generally consistent with
Hutchinson {\it et al.}\cite{hutchinson} for a spherical case
who give detailed $T$ dependence
of these quantities within the same approximation.
\begin{figure}
\hspace{0.5cm}
\epsfxsize=6cm
\epsfbox{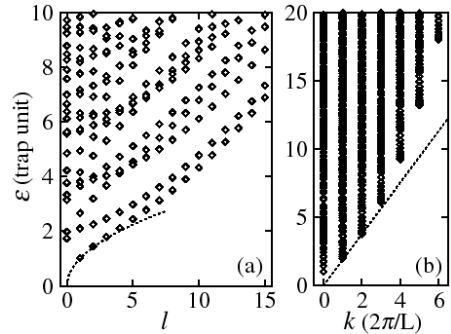}
\caption{
 Excitation spectra (a) along the angular momentum $l$ where the dotted 
line is the TFA prediction and (b) along the axial wave number $k$ where
the dotted line is the slope of the zero sound mode.
$n_z=2\times10^4/\mu {\rm m}$, $\omega_0=200{\rm Hz}$,
and $T=0.6\mu {\rm K}$.
}
\label{e:qt:qz}
\end{figure}

The whole excitation spectra characterized by the three quantum numbers
$q=(\kappa, l, k)$
are shown in Figs.~\ref{e:qt:qz}(a) and (b).
In Fig.~\ref{e:qt:qz}(a) the energy levels are plotted
as a function of the angular momentum $l$.
The behavior $\varepsilon=\omega_0{\sqrt l}$ expected
by hydrodynamic theory (or TFA) due to Stringari\cite{stringari} is
only approximately obeyed when $l$ is small in the present situation as shown by the dotted line.

The lowest three modes with $l=0,1,2$ for $\kappa=0$ are often discussed
theoretically and observed experimentally
on Rb case under a different geometrical confinement\cite{localized}.
The lowest dipole mode $l=1$ coincides with $\omega_0$,
almost independent of the interaction strength because
it corresponds to a rigid oscillation of the entire trapped gas.
Our result satisfies this criterion as inspected from Fig.~\ref{e:qt:qz}(a),
ensuring the accuracy of our numerical calculations.
Here we only remark that the relative position of the
monopole mode $\varepsilon(0,0,0)$  with $l=0$ and the quadrapole mode
$\varepsilon(0,2,0)$ with $l=2$ can be reversed when the area density along the $z$-axis changes 
where the energy eigenvalue is denoted
by $\varepsilon(\kappa,l,k)$ in terms of the above quantum numbers.

\begin{figure}
\hspace{1cm}
\epsfxsize=5cm
\epsfbox{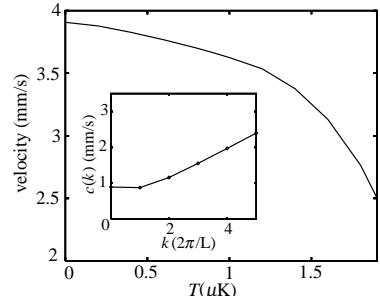}
\caption{
{} $T$ dependence of the sound velocity.
$n_z = 2\times 10^4/\mu {\rm m}$ and
$\omega_0 = 200{\rm Hz}$. 
The inset shows the full dispersion relation $c(k)$
of the zero sound mode
as a function of $k$ at $T=1.2\mu {\rm K}$.
}
\label{disper}
\end{figure}

The zero sound velocity $c$ is given by the slope of the lower edge of the spectrum along the $k$ direction at $k=0$ shown in Fig.~\ref{e:qt:qz}(b).
It is clear that $c(k)$ is dispersive from a linear to quadratic curve,
implying that the mode changes from the collective zero-sound to individual
particle excitations toward short wave numbers (Also see Fig.~\ref{disper}).
This crossover point corresponds to the inverse of the
characteristic coherent length $\xi$ of the system.
\begin{figure}
\hspace{1cm}
\epsfxsize=5cm
\epsfbox{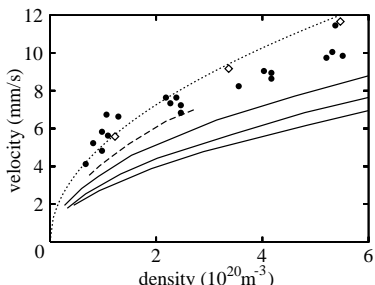}
\caption{
Zero sound velocity $c$ against the peak
condensate density $n_0(0)$ at low $T$.
The full curves are the harmonic potential case:
$n_z=2\times10^3$,
$1\times10^4$ and $2\times10^4/\mu {\rm m}$ from top to bottom.
The topmost one nearly corresponds to Zaremba's result.
The dashed line is the anharmonic case ($n_z=2\times10^3/\mu {\rm m}$).
The dots are the experimental data\protect\cite{andrews}.
The diamonds are the rigid wall potential case corresponding
to infinite homogeneous system (dotted line).
}
\label{sound:peak}
\end{figure}

In Fig.~\ref{sound:peak} we plot the sound
velocity $c$ as a function of the peak density
$n_0(0)$ of the condensate.
It is found that $c$ depends on the area density $n_z$
of the condensate along the axial direction, not only on $n_0(0)$.
This is physically reasonable because along each curve in Fig.~\ref{sound:peak}
the radial harmonic potential with $\omega_0$ varies,
which is implicit in this figure, keeping the area density $n_z$ fixed.
Therefore to fully characterize the system
we need the additional parameter $n_z$.
In other words, $c$ increases by increasing $\omega_0$ while fixing  $n_0(0)$.
These curves should be compared with that in the infinite homogeneous case.
The latter is calculated by the same scheme here except that
a rigid wall boundary condition is employed instead of a harmonic potential.
Our calculation indeed recovers the formula $c=\sqrt {n_0g/m}$
when the treated system is large enough
where the all relevant quantities are spatially uniform
except for the boundary\cite{isoshima}.
In this case there is no additional parameter like $n_z$.
The velocity $c$ increases as the area
density $n_z$ decreases as seen from Fig.~\ref{sound:peak}.
In the low area density limit $c$ tends to approach
Zaremba's result: $c=\sqrt {n_0g/2m}$.
Therefore, within the present situation of the radial harmonic confinement
our velocity never exceeds Zaremba's one.

In Fig.~\ref{disper} the $T$-dependence of $c$ is displayed.
As $T$ decreases $c(k=0)$ decreases,
signaling the loss of the rigidity of this macroscopic
quantum state toward the transition temperature.
In the inset of Fig.~\ref{disper} the full dispersion relation
 $c(k)={d\varepsilon\over dk}$
of the zero sound  mode is displayed.
The inverse of the crossover point from a linear to quadratic curve,
that is, the departure point from
a constant $c(k)$ gives rise to the coherent length
 $\xi$ ($\sim0.8\mu {\rm m}$ calculated for the present case).
It is known that the coherent length is given by
$\xi_0={1\over \sqrt{8\pi an_0}}$ for bulk homogeneous system.

\begin{figure}
\hspace{0.25cm}
\epsfxsize=7cm
\epsfbox{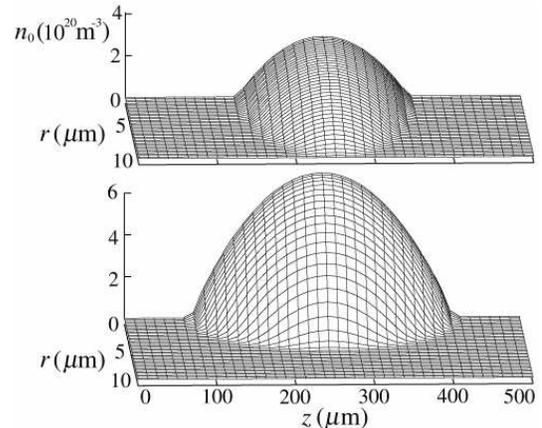}
\caption{
Stereographic view of the  condensate $n_0(r,z)$ at $T=0$ calculated by
Gross-Pitaevskii equation
at $\omega_0=200{\rm Hz}$ (upper) and $\omega_0=600{\rm Hz}$ (lower) 
under the axial potential fixed ($18{\rm Hz}$).
The total number is $5\times10^6$.
The area density $n_z$ at the center is seen
to be 4.2$\times10^4/\mu {\rm m}$ and
2.7$\times10^4/\mu {\rm m}$ respectively.
}
\label{3d}
\end{figure}

Let us now consider the experimental data for the zero sound 
by Andrews {\it et al.}\cite{andrews}.
For the reasonable area density $n_z\sim O(10^4)/\mu {\rm m}$ our
theoretical curves are far off the experimental data as
seen from Fig.~\ref{sound:peak}.
The experimental data are explained neither by our theoretical curves
nor by the Zaremba curve based on TFA\cite{zaremba,pethick}.
We argue the possible origins of this discrepancy:
(1) Uncertainty of the value of the repulsive
interaction $g$ or the s-wave scattering length $a$.
We have assumed $a=2.75{\rm nm}$ according to \cite{andrews}.
As inferred from the expression $c^2\propto a$ for infinite system,
$c$ is sensitive to the choice of the value $a$.
If we choose $a=4.9{\rm nm}$ as adopted before
by MIT group\cite{mit}, this discrepancy may be reduced.
(2) Anharmonicity of the trapping potential.
The possible anharmonicity such as $\tilde{k} r^4$ in the
trapping potential $V(r)$ may be present.
An example of our result for
typically $\tilde{k}=5\times 10^{-8} {\rm J/m^4}$ is shown
in Fig.~\ref{sound:peak} 
(dashed line),
giving rise to impressive improvement of our fitting.
Any anharmonicity of this type makes the condensate profile
a flat-topped bulky shape,
leading the system to more infinite homogeneous like.
This results in the more bulk like sound velocity.
Note that the reduction factor $\sqrt2$ in Zaremba's result
comes from a simple geometrical reason characteristic
to the parabolic shape condensate profile as mentioned before.
(3) As is seen from Figs.~\ref{sound:peak} and \ref{disper} the
zero sound
mode is dispersive.
The true zero sound measurement is only realized
when exciting long wave length limiting mode.
Any deviation from this limit tends to increase the estimate
for the velocity as seen from inset of Fig.~\ref{disper}.
This might be one reason why the low density data deviates even
from the infinite case, which gives the upper limit velocity.
Here we point out that since the low density case has smaller
aspect ratio, the sound propagation might not
be purely axial mode,
possibly exciting the radial mode simultaneously\cite{kurn}.
(4) The actual system is confined axially even though it is very long.
Thus along the axial direction the density is varying spatially,
this making experiment and theory for the zero sound
problematic.
We have performed the additional computation to see the shape and the
area density in the actual experimental
situations where the gas is trapped axially by a weak harmonic potential.
The Gross-Pitaevskii equation mentioned before is
solved under the axial trapping frequency $18{\rm Hz}$ 
fixed for various radial
trapping frequencies $\omega_0$ as in the actual experiment\cite{andrews}.
From Fig.~\ref{3d} which shows typical examples of the condensate
spatial shape, it is seen that upon varying $\omega_0$ not
only the peak density $n_0(0)$ but also the length and width
of the condensate change, resulting in the change of the area density $n_z$.
Generally, $n_z$ increases as $\omega_0$, or  $n_0(0)$ increase.
Thus in the actual experiment the velocity data are traversing 
the various curves in Fig.~\ref{sound:peak} toward
the upper ones as $n_0(0)$ increases.
We also note that according to \cite{andrews} the sound velocity
is position-dependent and increases near the  end of the cigar.
This is understandable because the ``effective" density consisting
of the actual density and the confining potential increases there,
making $c$ larger.

In conclusion, we have calculated the whole excitation spectrum of
a cylindrical symmetric Bose-Einstein gas trapped
by an axial harmonic potential within the Popov approximation
of the Bogoliubov framework.
The zero sound compressional mode is analyzed in detail.
The sound velocity and its temperature dependence are evaluated.
The recent experiment by Andrews {\it et al.}\cite{andrews} on Na gas
is discussed in light of the present theory,
pointing out several possible origins of the disagreement
between our theory and their result on the sound velocity.
It is our hope that the whole set of physical measurable quantities determined
here in a self-consistent manner further serves
to experimentally check the internal consistency of the Bogoliubov framework,
which has been remarkably successful so far.
We don't think the sound velocity an exception.

We thank D. M. Kurn, M.I.T. for helpful communication on their experiment
and also T. Yabusaki and Y. Takahashi, Kyoto Univ.
for educating us regarding basics of experimental methods.


\end{document}